\documentclass[%
 reprint,
 twocolumn,
 longbibliography,
 superscriptaddress,
 amsmath,amssymb,
 aps, physrev,
]{revtex4-2}
\usepackage{amsmath}
\usepackage{amsfonts}
\usepackage{amssymb}
\usepackage{graphicx}
\usepackage{verbatim}
\usepackage{makecell}
\usepackage{longtable}

\usepackage{siunitx}

\usepackage{hyperref}

\newcommand{\delay}{\tau_d}
\newcommand{\cyclecount}{k}
\newcommand{\activation}{\sigma}
\newcommand{\weightdecay}{\gamma}
\renewcommand{\omega}{\dot{\theta}_r}

\begin{document}
\title{Forward and reverse delay-driven hippocampal replay\\ without symmetric plasticity}

\author{Georg Reich}
\email[Contact author: ]{gr@ni.tu-berlin.de} 
\affiliation{Institute of Neuroinformatics, University of Zurich and ETH Zurich, Zurich, Switzerland}
\affiliation{Neural Information Processing Group, Technical University Berlin, Berlin, Germany}

\author{Matthew Cook}
\affiliation{Institute of Neuroinformatics, University of Zurich and ETH Zurich, Zurich, Switzerland}
\affiliation{University of Groningen, Groningen, The Netherlands}

\author{Klaus Obermayer}
\affiliation{Neural Information Processing Group, Technical University Berlin, Berlin, Germany}

\author{Pau Vilimelis Aceituno}
\affiliation{Institute of Neuroinformatics, University of Zurich and ETH Zurich, Zurich, Switzerland}

\date{\today}

\begin{abstract}
Hippocampal replay is a phenomenon observed in mammals and songbirds where neural activation sequences experienced during wakeful periods are repeated during rest or sleep. This mechanism is believed to play a crucial role in episodic memory consolidation, retrieval, and planning. Interestingly, replay can occur in both forward and reverse temporal orders and across a wide range of increased speeds.
The learning of activation sequences has traditionally been modeled by temporally asymmetric Hebbian plasticity rules, which explain replay as a chain of activity. The inability of such rules to strengthen backward connections has led to the widespread belief that they cannot account for bidirectional replay. Existing theoretical work therefore explains reverse replay through symmetric connections, which would be formed through symmetric plasticity rules. 
We propose a delay-coupled neural field model of CA3, where activity propagates as a wave in a ring, and replays occur both forwards and backwards at different speeds. Our model can learn bidirectional replay after a single noisy exposure to a stimulus with either symmetric or asymmetric plasticity. We derive a low-dimensional description of the dynamics and analyze the stability of possible replay speeds.
\end{abstract}
\maketitle

\section{Introduction}
After a single exposure to a novel experience, hippocampal and cortical circuits have been observed to repeat their activation sequences during rest and sleep in a phenomenon called replay~\cite{berners-lee_hippocampal_2022}. This mechanism is believed to play a crucial role in episodic memory consolidation, retrieval, and planning retrieval~\cite{chen_how_2023}. 
Despite only receiving stimuli at real-world speed in a predefined order, neural circuits in CA3 learn to replay sequences both forwards and in reverse~\cite{foster_reverse_2006, gupta_hippocampal_2010}. These replays also occur at various speeds, sometimes compressing the duration of the original events by up to 20 times~\cite{denovellis_hippocampal_2021}. Reverse replay is of particular interest because of its proposed role in value learning, as it is uniquely modulated by reward and occurs more frequently during wakefulness than during slow-wave sleep~\cite{ambrose_reverse_2016, findlay_evolving_2021}. 

Asymmetric learning rules provide a simple explanation for forward replay: synapses along the experienced firing sequence are strengthened and can then reactivate as a chain of activity \cite{fiete_spike-time-dependent_2010}. However, temporally anticausal connections get depressed, which means that such a chain cannot be activated in reverse. This has led to the widespread belief that an asymmetric learning mechanism cannot account for bidirectional replay~\cite{gupta_hippocampal_2010, haga_recurrent_2018, ecker_hippocampal_2022, milstein_offline_2023, mallory_time_2025}. 
An in vivo observation of symmetric STDP in mature CA3 synapses~\cite{mishra_symmetric_2016}, where backward connections in an activation sequence are equally potentiated, has so far been the primary experimental evidence for symmetric connectivity in recurrent hippocampal circuits~\cite{chenkov_memory_2017}. 
More recent evidence has found that CA3 plasticity is only symmetric in familiar environments but highly asymmetric during novel experiences~\cite{madar_btsp_2023}. Such novel experiences are thought to be most important for replay: a single traversal across a novel track is sufficient for replay to arise~\cite{foster_reverse_2006}, and the rate of replays declines with experience in a given environment~\cite{gupta_hippocampal_2010, foster_reverse_2006,findlay_evolving_2021}. Given this conflicting evidence regarding the symmetry of learning rules for replay, the mechanism behind reverse replay remains an open question. 
In this study, we introduce a possible mechanistic explanation for bidirectional replay, where neural activity is driven by its own delayed recurrent signal. Unlike previous explanations for reverse replay, our proposed mechanism does not rely on symmetric connectivity. We highlight this by modeling synaptic strengths using temporally asymmetric Hebbian learning. Our model predicts that possible replay speeds occur in discrete steps and become less stable with increasing speed. The speeds and their stability are determined by the stimulus and biophysical properties of the circuit, such as its firing rate time constant and transmission delays.

\section{Neural Field Model and Learning}
In this section, we introduce the neural field model during learning and obtain analytical solutions.

The idea of studying periodic neural fields with tools from harmonic analysis is inspired by \citeauthor{aceituno_resonances_2020}\cite{aceituno_resonances_2020, aceituno_insect_2023}, but we apply different techniques to handle the nonlinear neural activation function.

\subsection{Neural activity and plasticity model}\label{sec:model}
The firing rate \({ r(x, t) }\), representing the average number of action potentials per unit time at position \({ x }\) and time \({ t }\) in a one-dimensional hippocampal CA3 circuit, is modeled by
\begin{gather*}
    \tau_r {\partial_t} r(x, t) = -r(x, t) + \activation\left(u(x, t) + \left[\text{w} * r_{\text{pre}}\right](x, t)\right),\\
    \left[\text{w} * r_{\text{pre}}\right](x, t) := \int_0^T \text{w}(\Delta x, t) r(x - \Delta x, t - \delay) d\Delta x,
\end{gather*}
where \({ \activation }\) is a Heaviside step function representing the neural response, the kernel \({\text{w}(\Delta x, t)}\) represents the synaptic weight from position \({ x }\) to position \({ x + \Delta x }\) at time \({ t }\), and \({ u(x, t) }\) is the external input. The rate of change of \({ r(x, t) }\) is influenced by the effective time constant \({ \tau_r }\). The presynaptic activity \( {r_{\text{pre}}(x, t) := r(x, t-\delay)}\) with delay \({ \delay }\) represents the delayed firing rate of recurrent network interactions. Delays between action potentials at the somas of connected neurons are a result of axonal and dendritic conduction and synaptic transmission. 

Dynamical systems with delays are difficult to study unless we enforce periodicity, which allows us to treat a delay as a phase shift. Conveniently, such periodicity is found in grid cells in the entorhinal cortex~\cite{gardner_toroidal_2022}, which accounts for a significant portion of hippocampal inputs. We define the spatial domain of this external input \({ u(x, t) }\) as a circle, \({ \mathbb{R} / T\mathbb{Z} }\), where \({ T }\) is the period of the system, implying that \({ x = x + \cyclecount T }\) for any integer \({ \cyclecount }\). Furthermore, \({ u(x, t) }\) is periodic in the temporal domain, forming a traveling wave. We use arbitrary spatial units and set the phase velocity to 1 so that the wavelength is equal to the period \({ T }\). This results in symmetry between space and time, and simplifies the notation.
\begin{gather*}
  u(x, t) = u(x, t + T) = u(x + T, t), \\
  u(x, t) = u(0, t-x)
\end{gather*}
We model the evolution of synaptic weights through differential Hebbian learning, which is generally defined as \(\tau_{\text{w}}\partial_t\text{w} = r_{\text{pre}} \partial_tr\)~\cite{porr_isotropic_2003, zappacosta_general_2018}. Assuming separation of time scales \({ \tau_{\text{w}} \gg \tau_r }\), differential Hebb in the spatial continuum limit can be written as~(Appendix~\ref{sec:kernelrepresentation})
\begin{gather*}
\tau_{\text{w}}{\partial_t} \text{w}(\Delta x, t) = \frac{1}{T}\left[r_{\text{pre}} \boldsymbol{\star} {\partial_t} {r}\right](\Delta x, t) - \weightdecay \text{w}(\Delta x, t),\\
\left[r_{\text{pre}} \boldsymbol{\star} {\partial_t} {r}\right](\Delta x, t) := \int_0^T {r}(x', t-\delay){\partial_t} {r}(\Delta x + x', t)dx',
\end{gather*}
where \({ \tau_{\text{w}} }\) is the synaptic plasticity time constant, \({ \boldsymbol{\star} }\) is the cross-correlation operator, and \({ \weightdecay }\) is the homeostatic decay factor.

\subsection{Analytical firing rate solution}
We assume that in the input-driven regime, the input is much stronger than recurrent signals
\begin{equation}
  \label{eq:inputdriven}
  \left|u(x, t)\right| \gg \left|\left[\text{w} * r_{\text{pre}}\right](x, t)\right|,\\
\end{equation}
such that we can write the rate dynamics as
\begin{equation*}
\tau_r {\partial_t} r(x, t) = -r(x, t) + \activation\left(u(x, t)\right).
\end{equation*}
We note that the stimulus induces the same symmetry of time and space in neural activity.
For a large enough \({t}\), we thus have
\begin{gather}
  \label{eq:ratesymmetry}
  {r(x, t+t') = r(x-t', t)},\\
  \label{eq:ratediffsymmetry}
  {\partial_t} r(x, t) = -{\partial_x} r(x, t).
\end{gather}
By transforming the input-driven rate dynamics into the frequency domain and using the symmetry property~\eqref{eq:ratediffsymmetry}, we obtain the analytical solution~\eqref{eq:analyticalrate}. We write the spatial Fourier transform of a function \(f(x, t)\) as ${\mathcal{F}_x\left[f\right]\left(\xi, t\right)}$, which is sometimes abbreviated as \({\hat{f}}\).
\begin{align}
\mathcal{F}_x\left[\tau_r \partial_t r + r\right](\xi, t) &= \mathcal{F}_x\left[\activation(u(x, t))\right](\xi, t) \nonumber\\
(-\tau_r 2 \pi i \xi + 1)\hat{r}(\xi, t) &= \widehat{\activation(u)}(\xi, t) \nonumber\\
\label{eq:analyticalrate}
r(x, t) &= \mathcal{F}^{-1}_\xi\left[\frac{\widehat{\activation(u)}(\xi, t)}{-\tau_r 2 \pi i \xi + 1}\right](x, t)
\end{align}

\begin{figure*}[ht!]
  \includegraphics[width=0.85\linewidth]{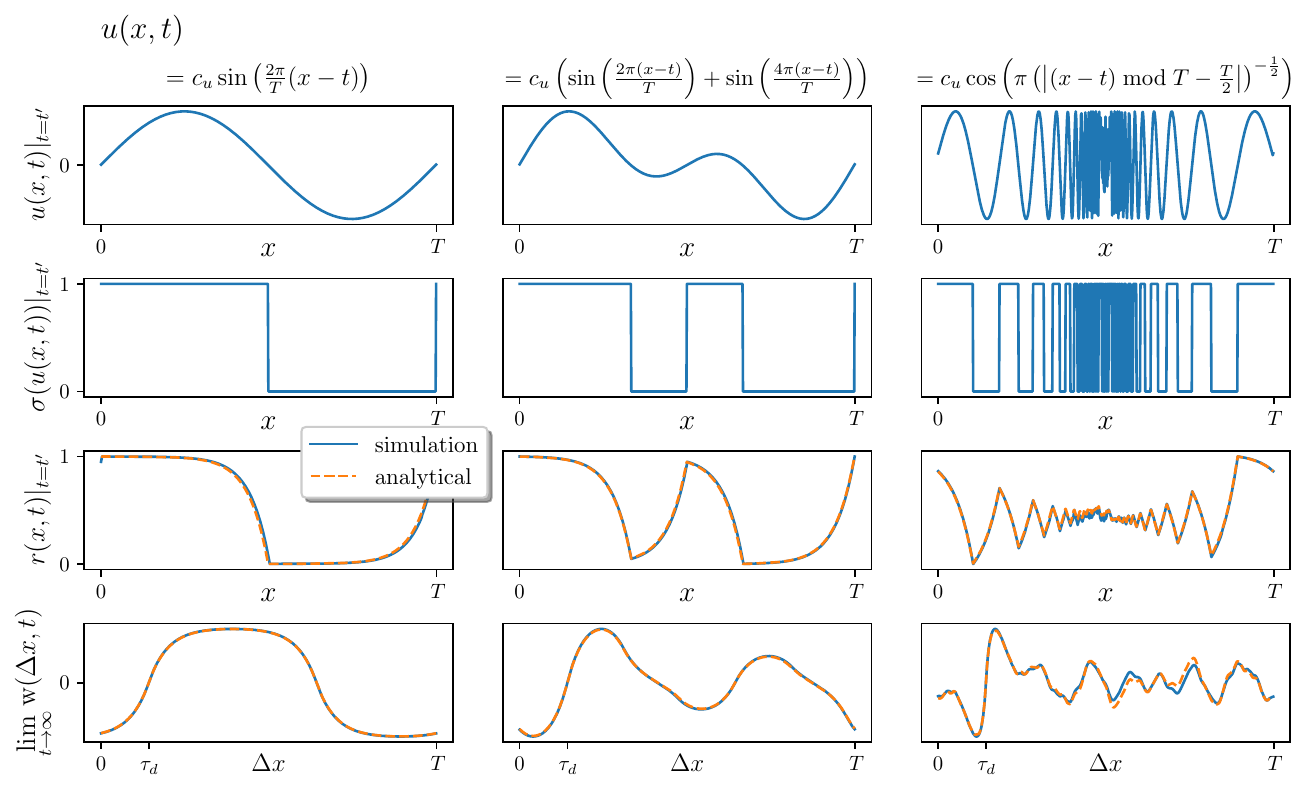}
  \vspace*{-0.3cm}
  \caption{Snapshots of external stimulation patterns with scale factor \(c_u\) (first row), neural response (second row), firing rate (third row), and synaptic weights after learning converges (last row) for different stimulation patterns (three columns). Analytical solutions for firing rate and weights at equilibrium are overlaid on top of values obtained from simulation.}
  \label{fig:input_driven_overview}
\end{figure*}

We observe from Eq.~\eqref{eq:analyticalrate} that the firing rate is just a low-pass filtered \({\activation(u(x, t))}\) with cutoff frequency \(\tau_r^{-1}\) (compare Fig.~\ref{fig:input_driven_overview}).

\subsection{Weight equilibrium}
When the synaptic weights are at equilibrium and the system no longer learns, the neural activity is periodic, and the weights remain constant.

\begin{figure}[ht!]
  \centering
  \includegraphics[width=0.78\linewidth]{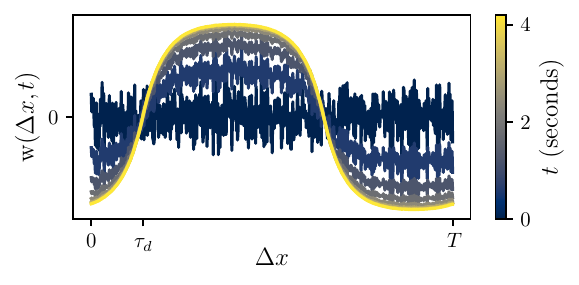}
  \vspace*{-0.2cm}
  \caption{Evolution of synaptic weights during stimulation. Given a random initialization, the weights learn the pattern and saturate after approximately 60 stimulation cycles for our chosen parameters.}
  \label{fig:weight_kernel_evolution}
\end{figure}
{
\allowdisplaybreaks
The learning dynamics converge to a fixed weight profile, which can be obtained by averaging the weight updates from Eq.~\eqref{eq:ratedynamics} over one period. 
\begin{align*}
0 &= \tau_{\text{w}}{\partial_t} \text{w}(\Delta x, t) \\
0 &= \frac{1}{T}\left[r_{\text{pre}} \boldsymbol{\star} {\partial_t} {r}\right](\Delta x, t) - \weightdecay \text{w}(\Delta x, t) \\
\intertext{Using the complex conjugate \(\overline{\cdot}\), and cross-correlation theorem \({\mathcal{F}[f \boldsymbol{\star} g] = \overline{\mathcal{F}[f]} \cdot \mathcal{F}[g]}\):}
0 &= \frac{1}{T}\mathcal{F}^{-1}_\xi\left[\overline{\widehat{r_{\text{pre}}}}\widehat{{\partial_t} {r}}\right](\Delta x, t) - \weightdecay \text{w}(\Delta x, t) \\
0 &= \overline{\widehat{r_{\text{pre}}}}\widehat{{\partial_t} {r}} - \weightdecay T \hat{\text{w}} \\
\intertext{Using the firing rate symmetries \eqref{eq:ratesymmetry} and \eqref{eq:ratediffsymmetry}:}
  &= \overline{e^{2\pi i \xi\delay} \hat{r}} (-2 \pi i \xi) \hat{r} - \weightdecay T \hat{\text{w}}\\
  &= -2 \pi i \xi {e^{-2\pi i \xi\delay}} {\left|\hat{r}\right|^2} - \weightdecay T \hat{\text{w}}
\end{align*}
}
The energy spectral density \({\left|\hat{r}\right|^2}\) can be related to the power spectral density \({S_x\left[r\right](\xi, t)}\) via the Wiener-Khinchin theorem.
Solving for \({\text{w}(\Delta x, t)}\), we obtain the analytical steady-state solution for the synaptic weights
\begin{equation}
\label{eq:analyticalweights}
\text{w}(\Delta x, t) = \mathcal{F}^{-1}_\xi\left[\frac{1}{\weightdecay T} (-2 \pi i \xi) {e^{-2\pi i \xi\delay}} S_x\left[r\right]\right](\Delta x, t).
\end{equation}
This solution can be understood by interpreting the expression in the frequency domain as a filter applied to the power spectrum. The factor $-2\pi i \xi$ corresponds to the derivative $\partial_{t}$ in the differential Hebbian learning rule. Multiplication with \({\xi}\) implies that there is no constant offset term, which results in a balance between inhibition and excitation. This kernel is then spatially shifted by \(\delay\), as dictated by \(e^{-2\pi i \xi\delay}\). This spatial shift can be observed in Fig.~\ref{fig:weight_kernel_evolution}, where we show snapshots of the weight kernel $\text{w}$ during learning for \SI{4}{\second}. Given the scaling by \(\frac{1}{\weightdecay T}\), synaptic weights are weaker and converge faster for a stronger decay factor \({\weightdecay}\). 
In Fig.~\ref{fig:input_driven_overview}, we confirm Eqs.~\eqref{eq:analyticalrate} and \eqref{eq:analyticalweights} in simulation for periodic traveling waves of different shapes. 

\section{Replay Solutions}
\begin{figure}[ht!]
  \includegraphics[width=0.75\linewidth]{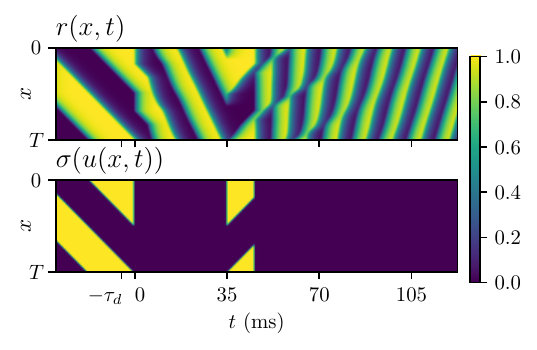}
  \vspace*{-0.3cm}
  \caption{The top plot depicts neural activity of our model over time, when stimulated with the pattern of the bottom plot after learning until convergence. After the stimulus is turned off at $t=0$, the model replays the pattern at approximately double the speed. After the time-reversed original stimulus is given for around \SI{10}{{\milli\second}}, the model replays in reverse at around three times the speed.}
  \label{fig:basic_replay}
\end{figure}

We observe that when external stimulation is turned off after learning, our model repeats the periodic pattern at an increased speed (Fig.~\ref{fig:basic_replay}). Interestingly, replay changes the direction and further increases the speed after we present the model with a stimulus in the opposite direction. To study this behavior analytically, we only consider the case of a single traveling bump given by the fundamental mode $\xi = 1/T$. We can express it in polar form
\begin{equation}
  \hat{r}\left(\xi=\frac{1}{T}, t\right) = a_r(t) e^{i \theta_r(t)},
  \label{eq:lowdim_decomposition}
\end{equation}
where $a_r(t) = |\hat{r}(1/T, t)|$ and $\theta_r(t) = \arg(\hat{r}(1/T, t))$ are the instantaneous amplitude and phase of the activity bump, respectively.
During replay, the only input is the recurrent synaptic current:
\begin{equation}
\tau_r \frac{\partial r(x, t)}{\partial t} = -r(x, t) + \sigma\left(\left[\text{w} * r_{\text{pre}}\right](x, t)\right).
\label{eq:replay_pde}
\end{equation}
The low-dimensional dynamics are then given by
\begin{equation}
    \label{eq:lowdim_dynamics_intro}
    \tau_r\frac{d}{dt} \hat{r}\left(\frac{1}{T}, t\right) = - \hat{r}\left(\frac{1}{T}, t\right) + \widehat{\sigma(\text{w}*r_{\text{pre}})}\left(\frac{1}{T}, t\right).
\end{equation}
Because the self-propagating recurrent input \( \activation(\cdot) \) takes the form of a periodic square wave, the fundamental Fourier component of \( \activation(\text{w} * r_{\text{pre}}) \) has a fixed amplitude \( 2/\pi \), independent of \( a_r(t-\delay) \).
The phase of the square wave is determined by the zero-crossings of the convolution. We approximate this phase as the fundamental phase of the convolution plus a correction parameter \(\varepsilon\) that accounts for any skewness in the activity profile \(r\). Thus, the Fourier transform of the recurrent signal evaluates to
\begin{equation}
  \widehat{\sigma(\text{w}*r_{\text{pre}})}\left(\frac{1}{T}, t\right) \approx \frac{2}{\pi} e^{i \left(\theta_{\text{w}}(t) + \theta_r(t-\delay) + \varepsilon\right)}.
  \label{eq:fourier_H}
\end{equation}

Using the analytical solution given by Eq.~\eqref{eq:analyticalweights} with ${\xi=\frac{1}{T}}$, the fundamental phase of learned synaptic weights becomes
\begin{align}
  \theta_{\text{w}}(t)&= \text{arg}\left(\hat{\text{w}}\left(\xi=\frac{1}{T}, t\right)\right)\nonumber\\
  &= \text{arg}\left(\frac{1}{\weightdecay T^2} (-2 \pi i){e^{-i\left(\frac{1}{T}2\pi\delay\right)}} S_x[r]\left(\frac{1}{T}\right)\right)\nonumber\\
  &= \text{arg}(-i) - 2\pi\frac{\delay}{T} + \text{arg}\left(\hat{r}\right) -\text{arg}\left({\hat{r}}\right)\nonumber\\
  \label{eq:weightsphase}
  &= -\frac{\pi}{2} - 2\pi\frac{\delay}{T},
\end{align}
because the power spectral density $S_x[r]$ is real and positive. We assume \({\tau_{\text{w}} = \infty}\) and can therefore denote the phase of learned weights as constant, \(\theta_{\text{w}}(t) = \theta_{\text{w}}\).

\begin{figure}[!htbp]
  \centering
  \includegraphics[width=0.9\linewidth]{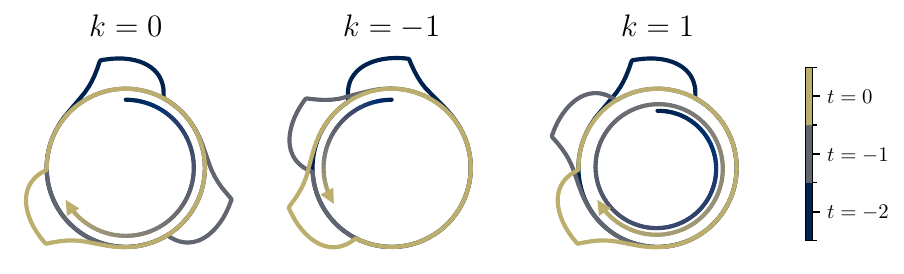}
  \vspace*{-0.3cm}
  \caption{Illustration of three traveling waves of neural activity taking different paths on the circle for the same distance between starting and end point.}
  \label{fig:introduction_illustration}
\end{figure}

Substituting the polar form \eqref{eq:lowdim_decomposition} and the approximation \eqref{eq:fourier_H} into the projected dynamical equation \eqref{eq:lowdim_dynamics_intro} yields the complex-valued ODE
\begin{equation*}
  \tau_r\frac{d}{dt} \left[a_r(t) e^{i \theta_r(t)}\right] = -a_r(t) e^{i \theta_r(t)} + \frac{2}{\pi} e^{i (\theta_{\text{w}} + \theta_r(t-\delay) + \varepsilon)}.
\end{equation*}
Expanding the derivative and dividing by $e^{i\theta_r(t)}$ yields
\begin{align}
  &\tau_r\left(\dot{a}_r(t) + i a_r(t)\dot{\theta}_r(t)\right)\nonumber\\
  &= -a_r(t) + \frac{2}{\pi} e^{i\left( \theta_{\text{w}} + \theta_r(t-\delay) - \theta_r(t) + \varepsilon\right)}.
\end{align}
Separating the real and imaginary parts gives us the final evolution equations for the amplitude and phase:
\begin{subequations}\label{eq:reduced_system}
\begin{align}
  &\tau_r \dot{a}_r(t) = -a_r(t) + \frac{2}{\pi} \cos(\theta_{\text{w}} + \theta_r(t-\delay) - \theta_r(t) + \varepsilon), \label{eq:amp_eq} \\
  &\tau_r a_r(t) \dot{\theta}_r(t) = \frac{2}{\pi} \sin(\theta_{\text{w}} + \theta_r(t-\delay) - \theta_r(t) + \varepsilon). \label{eq:phase_eq}
\end{align}
\end{subequations}
Because the spatial Fourier phase of a right-traveling wave rotates clockwise, its time derivative $\dot{\theta}_r(t)$ is negative, but corresponds to a positive phase velocity
\begin{equation}
    v(t) = -\frac{T}{2 \pi} \omega(t),
\end{equation}
which was $v=1$ during stimulus presentation (Eq.~\ref{eq:ratediffsymmetry}). We seek self-consistent solutions in which the amplitude and phase velocity are constant, that is, \(\dot{a}_r(t) = 0\) and \(v(t) = v^*\). For instance, replay modes that are twice as fast as the original stimulus have \(v^* = 2\) and reverse replay modes have a negative~\(v^*\).
Our approach to find such solutions \((a_r^*, \omega^*)\) is to perform the substitution
\begin{align}
  &\theta_r(t-\delay) - \theta_r(t) = -\int_{t-\delay}^{t} \dot{\theta}_r(t') dt' \nonumber\\
  \label{eq:stabletravelingwave}
  &= \delay \omega^* + 2\pi \cyclecount, \quad \cyclecount \in \mathbb{Z}.
\end{align}
where \(\cyclecount\) keeps the phase difference in \([0, 2\pi)\) and represents the cycle count of the activity within the delay. A cycle count of \(\cyclecount=0\) corresponds to a direct path between \(\theta_r(t-\delay)\) and \(\theta_r(t)\), a cycle count of \(\cyclecount=1\) corresponds to one additional cycle, and \(\cyclecount=-1\) to an additional cycle in the opposite direction (compare Fig.~\ref{fig:introduction_illustration}).
This leads to the following fixed-point conditions for a specific replay solution \( (a_r^*, \omega^*) \) and its corresponding phase correction~\( \varepsilon^* \):
\begin{gather}
    a_r^* = \frac{2}{\pi}\cos\left( \theta_{\text{w}} + \delay\omega^* + 2\pi\cyclecount + \varepsilon^*\right) \label{eq:fixed_amp_solution} \\
    -\tau_r a_r^*\omega^* = \frac{2}{\pi}\sin\left(\theta_{\text{w}} + \delay\omega^* + 2\pi\cyclecount + \varepsilon^*\right)\nonumber \\
    \tau_r \omega^* = \tan\left(-\theta_{\text{w}} - \delay\omega^* - 2\pi\cyclecount - \varepsilon^* \right) \label{eq:fixed_phase_velocity}
\end{gather}
Eq.~\eqref{eq:fixed_phase_velocity} defines the possible fixed replay speeds \( \omega^* \) for different cycle counts \(\cyclecount\) and can be solved numerically. For replay of fixed speed \(\omega^*\), the optimal phase correction \(\varepsilon^*\) is a function of \(\omega^*\tau_r\). To provide an upper and lower bound for \(\omega^*\), we assume a given \(\varepsilon\) and perform a first-and third-order (Appendix~\ref{sec:cubicapproximation}) Taylor approximation of the inverse tangent. For realistic parameter values (see Appendix~\ref{sec:parametervalues}), this bound is tight for small replay speeds, which aren't affected by the low-pass filtering of activity propagation,
\begin{align}
  \tau_r\omega^* &\approx -\theta_{\text{w}} - \delay \omega^* - 2\pi \cyclecount - \varepsilon \nonumber\\
  \omega^* &\approx \frac{-\theta_{\text{w}} - 2\pi \cyclecount - \varepsilon }{\delay + \tau_r}\label{eq:linearapprox_weightphase}.
\end{align}
Plugging in Eq.~\eqref{eq:weightsphase}:
\begin{align}
  \label{eq:linearapprox}
  \omega^* &\approx \frac{2\pi(\cyclecount + \frac{1}{4} + \frac{\delay}{T}) - \varepsilon}{\delay + \tau_r}\\
  v^* &\approx \frac{T(\cyclecount + \frac{1}{4}) + \delay - \varepsilon}{\delay + \tau_r}.
\end{align}
From this equation, we see that replay velocity solutions become denser for higher values of \(\frac{{\delay + \tau_r}}{T}\), and the offset is determined by \(\delay\).

\section{Replay Stability Analysis}
\begin{figure}[ht!]
  \includegraphics[width=0.75\linewidth]{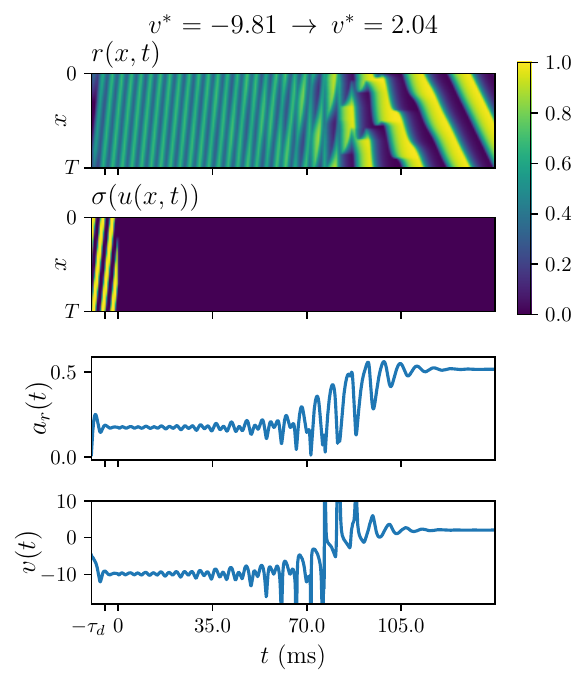}
  \vspace*{-0.3cm}
  \caption{We determine replay modes and initialize the model at a higher \(\left|v^*\right|\), with \(k=2\). After maintaining this higher speed for $\sim$\SI{50}{\milli\second}, it transitions to a lower speed (\(k=0\)). The two lower plots show how the replay mode instability leads to exponentially growing oscillations of amplitude and instantaneous replay speed until the system transitions to a stable solution. This transition from a higher to a lower speed is not a result of learning dynamics, as synaptic plasticity was turned off during replay.}
  \label{fig:transient_replay_speed}
\end{figure}
\begin{figure}[ht!]
  \centering
  \includegraphics[width=\linewidth]{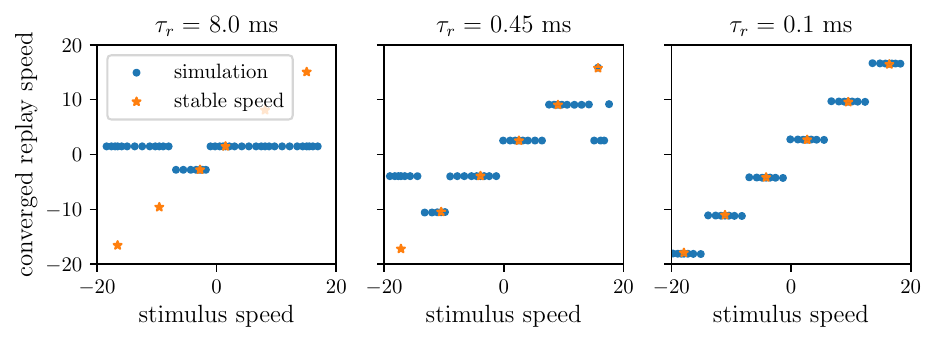}
  \vspace*{-0.5cm}
  \caption{By varying the speed of the initial stimulus, we can see the basins of attraction of different replay modes \(v^*\) depending on the firing rate time constant \(\tau_r\). On the diagonal, the orange stars denote solutions to Eq.~\eqref{eq:fixed_phase_velocity}, with \(k\) between \(-3\) and \(2\). As \(\tau_r\) decreases, more replay modes become stable.}
  \label{fig:converged_replay_speeds}
\end{figure}
The fixed replay speeds \(\omega^*\) identified by Eq.~\eqref{eq:fixed_phase_velocity} represent idealized, noiseless solutions. However, neural activity is inherently stochastic, and replay dynamics can be influenced by perturbations. As observed empirically (Fig.~\ref{fig:transient_replay_speed} and Fig.~\ref{fig:converged_replay_speeds}), simulations initialized at higher \(|\omega^*|\) will transition to a more robust mode with a lower \(|\omega^*|\) (typically where \(\cyclecount = 0\) or \(\cyclecount = 1\)). To understand this phenomenon and the robustness of different replay modes, we now analyze their linear stability.

The analysis is most transparent in a coordinate frame where the spatial phase of a stable traveling wave is stationary. Since the phase rotates at $-\omega^*$, we define a relative phase
\begin{equation}
\vartheta(t) = \theta_r(t) + \omega^* t. \label{eq:vartheta}
\end{equation}
In this frame, a stable replay solution corresponds to a fixed point \((a_r^*, \vartheta^*)\), where \(\vartheta(t) = \vartheta(t-\delay) = \vartheta^*\). Due to rotational symmetry, the specific value of \(\vartheta^*\) is arbitrary. By substituting this coordinate transformation into Eqs.~\eqref{eq:reduced_system} and defining a constant phase shift $\phi^* = \theta_{\text{w}} + \omega^*\delay + 2\pi\cyclecount + \varepsilon^*$ that absorbs the time delay (see Appendix~\ref{sec:rotating_coordinate_frame_derivation} for the full derivation), the system can be rewritten as 
\begin{subequations}\label{eq:rotating_frame_system}
\begin{align}
    &\tau_r \dot{a}_r(t) = -a_r(t) + \frac{2}{\pi} \cos\left(\phi^* + \vartheta(t-\delay) - \vartheta(t)\right), \label{eq:amp_dot_rotating_frame} \\
    &\tau_r \dot{\vartheta}(t) = \frac{1}{a_r(t)} \frac{2}{\pi} \sin\left(\phi^* + \vartheta(t-\delay) - \vartheta(t)\right) + \tau_r \omega^*.\label{eq:vartheta_dot_rotating_frame}
\end{align}
\end{subequations}

To probe the stability of the fixed point $(a_r^*, \vartheta^*)$, we introduce an infinitesimal perturbation vector \({\mathbf{p}(t) = [\delta a_r(t), \delta \vartheta(t)]^\top}\), such that
\begin{align}
    a_r(t) &= a_r^* + \delta a_r(t), \\
    \vartheta(t) &= \vartheta^* + \delta \vartheta(t).
\end{align}
Linearizing Eqs.~\eqref{eq:amp_dot_rotating_frame} and \eqref{eq:vartheta_dot_rotating_frame} with respect to these perturbations gives a system of linear delay-differential equations
\begin{equation}
    \tau_r \frac{d}{dt} \mathbf{p}(t) = \mathbf{M}_0 \mathbf{p}(t) + \mathbf{M}_{\delay} \mathbf{p}(t-\delay), \label{eq:linear_matrix_form}
\end{equation}
where the Jacobian matrices $\mathbf{M}_0$ and $\mathbf{M}_{\delay}$ contain the partial derivatives of the system with respect to the instantaneous and delayed state variables, evaluated at the fixed point. The detailed derivation of their elements is presented in Appendix~\ref{sec:stability_derivation}. The resulting matrices are
\begin{align}
    \mathbf{M}_0 = \begin{pmatrix} -1 & a_r^* \tau_r \omega^* \\ -\frac{\tau_r \omega^*}{a_r^*} & -1 \end{pmatrix}, \quad
    \mathbf{M}_{\delay} = \begin{pmatrix} 0 & -a_r^* \tau_r \omega^* \\ 0 & 1 \end{pmatrix}.
\end{align}
To find the characteristic equation for the stability eigenvalues $\lambda$, we assume a solution of the form $\mathbf{p}(t) = \mathbf{p}_0 e^{\lambda t}$. Substituting this Ansatz into Eq.~\eqref{eq:linear_matrix_form} yields
\begin{equation}
    \tau_r \lambda \mathbf{p}_0 e^{\lambda t} = \mathbf{M}_0 \mathbf{p}_0 e^{\lambda t} + \mathbf{M}_{\delay} \mathbf{p}_0 e^{\lambda (t-\delay)},
\end{equation}
which simplifies to the algebraic eigenvalue problem
\begin{equation}
    \left( \tau_r \lambda \mathbf{I} - \mathbf{M}_0 - \mathbf{M}_{\delay} e^{-\lambda \delay} \right) \mathbf{p}_0 = \mathbf{0}, \label{eq:algebraic_eigenproblem}
\end{equation}
where $\mathbf{I}$ is the $2 \times 2$ identity matrix. For non-trivial solutions $\mathbf{p}_0 \neq \mathbf{0}$, the determinant of the matrix in parentheses must be zero:
\begin{align*}
    &\det\left( \tau_r \lambda \mathbf{I} - \mathbf{M}_0 - \mathbf{M}_{\delay} e^{-\lambda \delay} \right) = 0\\ \label{eq:char_det_condition_matrix}
    &\det\begin{pmatrix} \tau_r \lambda + 1 & a_r^* \tau_r \omega^* (1 - e^{-\lambda \delay}) \\ -\frac{\tau_r \omega^*}{a_r^*} & \tau_r \lambda + 1 - e^{-\lambda \delay} \end{pmatrix} = 0\\
    &(\tau_r \lambda + 1)(\tau_r \lambda + 1 - e^{-\lambda \delay}) \\
    &\quad+ \left(a_r^* \tau_r \omega^* (1 - e^{-\lambda \delay})\right) \left(\frac{\tau_r \omega^*}{a_r^*}\right) = 0\\
    &(\tau_r \lambda + 1)^2 - (\tau_r \lambda + 1)e^{-\lambda \delay} + (\tau_r \omega^*)^2 (1 - e^{-\lambda \delay}) = 0
\end{align*}
This gives us the characteristic equation
\begin{equation}
    (\tau_r \lambda + 1)^2 + (\tau_r \omega^*)^2 = \left[(\tau_r \lambda + 1) + (\tau_r \omega^*)^2\right] e^{-\lambda \delay}. \label{eq:characteristic_equation}
\end{equation}
The stability of the fixed point $(a_r^*, \omega^*)$ is determined by the roots $\lambda$ of this transcendental equation. The fixed point is stable if all roots have $\text{Re}(\lambda) < 0$, and unstable if any root has $\text{Re}(\lambda) > 0$. A root with $\text{Re}(\lambda) = 0$ signals a potential bifurcation.

\begin{figure}[ht!]
  \centering
  \includegraphics[width=0.95\linewidth]{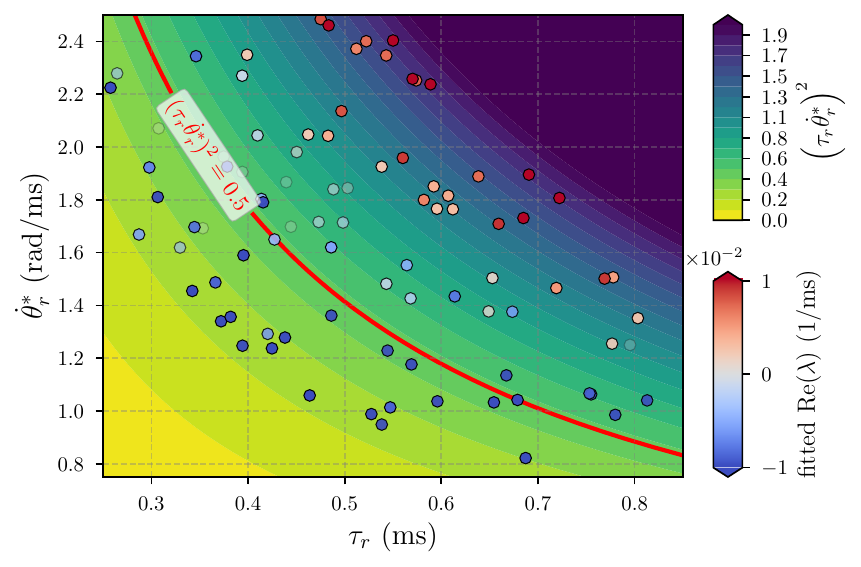}
  \vspace*{-0.3cm}
  \caption{Numerically determined stability of replay solutions compared to analytical lower bound of Hopf bifurcations. The background shows $(\tau_r\omega^*)^2$. Values below the red line ($(\tau_r\omega^*)^2=0.5$) indicate solutions that are analytically stable against Hopf instabilities. Scatter points are individual replay solutions (obtained by varying $\tau_r$, delay $\delay$, and cycle count $\cyclecount$), colored by their numerically fitted perturbation growth rate $\mathrm{Re}(\lambda)$ (blue: stable, red: unstable). Opacity reflects the fit's $R^2$. Unstable solutions are confined to regions where ${(\tau_r\omega^*)^2 > 0.5}$.}
  \label{fig:bifurcation_diagram}
\end{figure}

First, we note that due to the rotational symmetry of the phase $\vartheta$, Eq.~\eqref{eq:characteristic_equation} always has a root at $\lambda=0$, corresponding to a neutral (Goldstone) mode. However, this neutral stability does not explain the observed decay of high-speed solutions. A true instability, where the amplitude of the replay wave decays or grows uncontrollably, must arise from an eigenvalue acquiring a positive real part, typically through a Hopf bifurcation where a complex-conjugate pair of eigenvalues crosses the imaginary axis. We therefore seek the conditions for a Hopf bifurcation by setting $\lambda = i\Omega$ in Eq.~\eqref{eq:characteristic_equation}. For a solution to exist with real frequency $\Omega > 0$, the squared magnitudes of both sides of the equation must be equal. This requirement leads to an expression for the Hopf frequency (see Appendix~\ref{sec:Hopf_derivation} for details):
\begin{equation}
    \tau_r^2\Omega^2 = 2(\tau_r \omega^*)^2 - 1. \label{eq:Hopf_frequency_sq}
\end{equation}
For a Hopf bifurcation to occur with a non-zero frequency $\Omega$, we must have $\Omega^2 > 0$. From Eq.~\eqref{eq:Hopf_frequency_sq}, this restricts the possibility of such a bifurcation to the regime:
\begin{equation}
    2(\tau_r\omega^*)^2 - 1 > 0 \quad \implies \quad (\tau_r\omega^*)^2 > \frac{1}{2}. \label{eq:Hopf_condition_omega_tau}
\end{equation}
As detailed in Appendix~\ref{sec:Hopf_derivation}, evaluating the system below this boundary confirms that it is linearly stable. Consequently, $(\tau_r\omega^*)^2 > 1/2$ represents a necessary lower bound for the onset of instability. This condition can also be expressed as \({\tan^2(\phi^*) > 1/2}\).

To numerically assess the stability of simulated replay solutions, we estimate the real part of the dominant eigenvalue, $\mathrm{Re}(\lambda)$, associated with the linear stability Ansatz [Eq.~\eqref{eq:algebraic_eigenproblem}]. This is achieved by fitting an exponential trend to the envelope of the simulated amplitude $a_r(t)$. As shown in Fig.~\ref{fig:bifurcation_diagram}, empirically unstable replay modes (red points) are confined to the region where $(\tau_r\omega^*)^2 > 1/2$.

\section{Robustness}
In the previous sections, we studied the same neural field model with the parameter values given in Appendix~\ref{sec:parametervalues}. In this section, we study the requirements and limits of replay in our model by introducing variations.

\subsection{Stimulus variations}
\begin{figure}[ht!]
  \includegraphics[width=\linewidth]{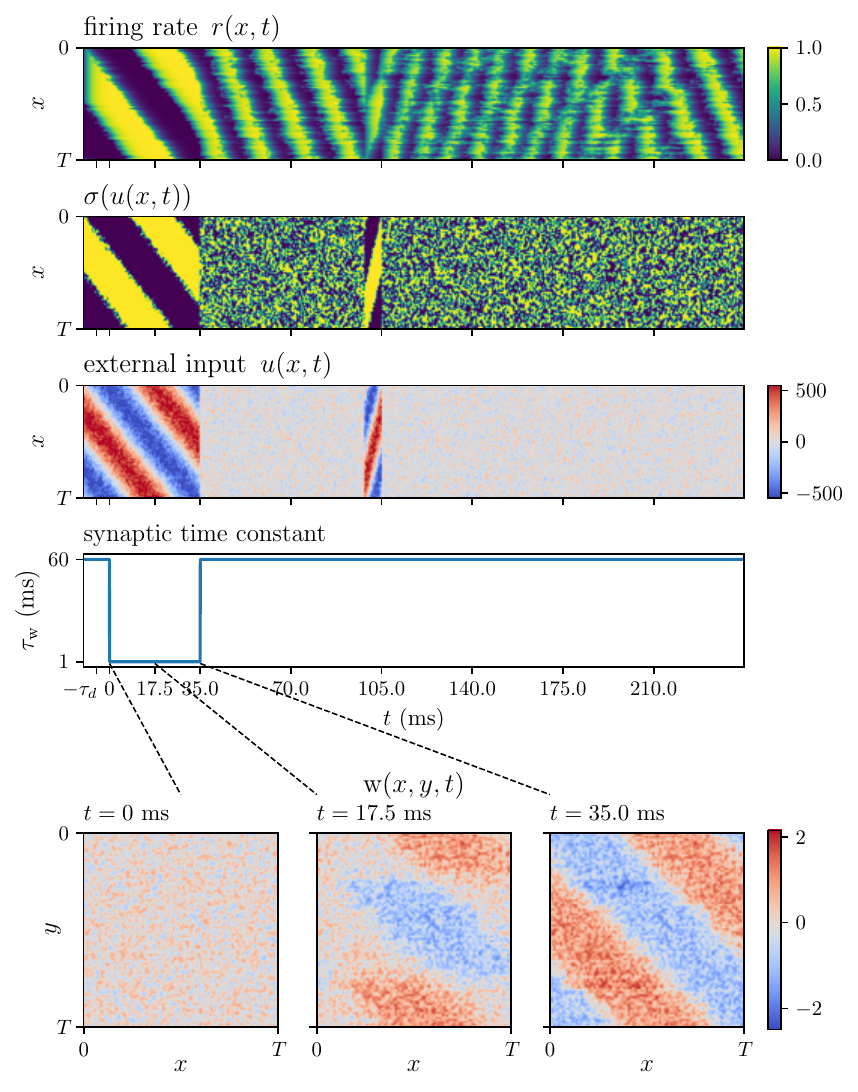}
  \caption{
    In this simulation, we introduce noise to the external input and only stimulate for one cycle (second and third subplot). The model successfully replays, and switches direction after being shown the reverse stimulus for \SI{6}{{\milli\second}}. The bottom subplot shows the evolution of the weight matrix during stimulus presentation, where we consider neuromodulation to decrease the synaptic plasticity time constant. Despite being in an analytically stable reverse replay mode, the model switches back to the slower forward replay after around \SI{80}{{\milli\second}} due to noise and the non-converged weights.
  }  \label{fig:realistic_replay}
\end{figure}
As an initial assessment of robustness, we only show the stimulus once and add coarse-grained noise (Fig~\ref{fig:realistic_replay}). To enable such "one-shot" learning, we temporarily reduce the synaptic plasticity time constant \(\tau_{\text{w}}\) during stimulus presentation, emulating neuromodulation from a novel experience. Since this violates the adiabatic assumption \({ \tau_{\text{w}} \gg \tau_r }\), we simulate the entire weight matrix instead of reducing it to a convolution kernel. Despite the noise, the model can already perform replay in both directions after a single experience of the stimulus. Due to the external noise and non-converged weights, the model switches back to forward replay after some time.

\begin{figure}[ht!]
  \centering
  \includegraphics[width=0.75\linewidth]{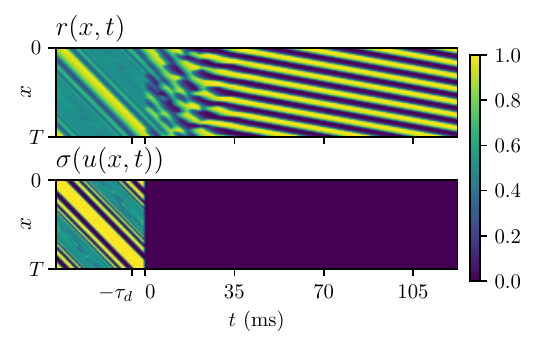}
  \vspace*{-0.3cm}
  \caption{After training and initializing a model with the third pattern of Fig.~\ref{fig:input_driven_overview}, it replays a pattern with multiple regularly spaced bumps.}
  \label{fig:chirp_replay}
\end{figure}
Next, we investigated if replay is possible with a traveling circular chirp stimulus (Fig.~\ref{fig:chirp_replay}). Interestingly, the model replayed a traveling wave pattern with multiple evenly spaced bumps. Since the learned weights contain a low-pass filter (compare Fig.~\ref{fig:input_driven_overview} bottom right), the lower frequencies of the chirp dictate the spatial spacing between replay bumps.

\subsection{Generalization to Fitzhugh-Nagumo dynamics}
\begin{figure}[ht!]
  \centering
  \includegraphics[width=0.85\linewidth]{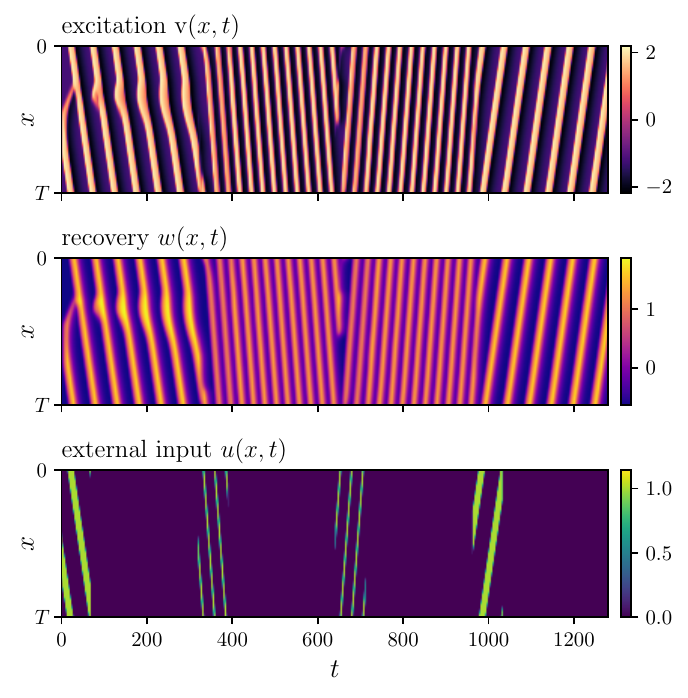}
  \vspace*{-0.3cm}
  \caption{Velocity-multiplexed traveling pulses in the time-delayed FitzHugh-Nagumo model~\eqref{eq:fhn_dynamics}.
    Top: excitation field $\text{v}(x, t)$, showing four successive write-hold cycles at different quantized velocities.
    Middle: the recovery variable $w(x, t)$, which trails the excitation front and enforces pulse termination.
    Bottom: external forcing $u(x, t)$ applied during each write window. During each hold phase the pulse train continues to propagate at the written speed without external drive. Parameters are given in Appendix~\ref{sec:fhn_parameters}.
  }
  \label{fig:fhn}
\end{figure}
To confirm that the replay speed quantization is not a quirk of our specific neural field formulation, we investigate the phenomenon in a qualitatively different setting: the diffusely coupled FitzHugh-Nagumo (FHN) model~\cite{fitzhugh_impulses_1961, nagumo_active_1962}, where we additionally introduce delayed self-coupling. It consists of the excitation field $\text{v}(x, t)$ and recovery field $w(x, t)$ on a periodic domain, governed by
\begin{subequations}\label{eq:fhn_dynamics}
\begin{gather}
  \tau_\text{v}\partial_t \text{v} = \partial_x^2 \text{v}
    + \text{v} - \frac{\text{v}^3}{3} - w
    + \sigma\left(\text{v}(x, t{-}\delay)\right)
    + u(x, t),\label{eq:fhn_v}\\
  \tau_w\partial_t w = \text{v} + a - b w, \label{eq:fhn_w}
\end{gather}
\end{subequations}
where $\tau_\text{v}$ and $\tau_w \gg \tau_\text{v}$ are the excitation and recovery time constants, $\partial_x^2 \text{v}$ is the spatial diffusion coupling, and the parameters $a$ and $b$ are chosen such that neural activity vanishes quickly without additional input. Sustained propagating activity is enabled by delayed self-coupling $\sigma(\text{v}(x, t{-}\delay))$, which uses the same Heaviside nonlinearity as our original neural field, but doesn't contain a synaptic weight convolution. It is therefore entirely local and requires no learning. The recovery variable $w$ provides a slow negative feedback mechanism and shapes the traveling pulse.

We determined possible replay speeds numerically and verified them in simulation (Fig.~\ref{fig:fhn}), showing that our mechanism for bidirectional replay is generalizable to excitable systems without non-local coupling. Since this model doesn't have asymmetric coupling, replay speeds are symmetric around zero.

\section{Discussion}
We have introduced a novel mechanism for bidirectional and multi-speed replay in recurrent neural circuits, driven by the interaction of neural activity with its own delayed recurrent signal. Our central contribution is the demonstration that such bidirectional replay can emerge from temporally asymmetric Hebbian learning, a finding that challenges the widespread belief that symmetric connectivity is necessary for reverse replay. In fact, our model also works with symmetric connectivity, only requires a single exposure to the stimulus, and is robust to noise. 
By modeling the system as a neural field with periodic boundary conditions, we derived analytical solutions for the learned synaptic weights and the discrete set of possible replay velocities. A linear stability analysis of these solutions revealed that replay stability is inversely related to its speed and firing rate time constant. We derived a lower bound of $(\tau_r\omega^*)^2 > 1/2$ for Hopf bifurcations and validated this prediction in simulations. In this section, we take a critical look at the neuroscientific relevance of our contribution.

\subsection{Relation to previous work}
Although at first glance our mechanism might seem similar to the continuous wagon wheel illusion, where a forward spinning spoked wheel is perceived as spinning backwards~\cite{vanrullen_continuous_2006}, it is qualitatively different: The wave travel direction in our mechanism is not an illusion, and does not require an observer who perceives information in discrete time steps. Delays are necessary for our system, and we couldn't find any existing work with a delay-differential equation like Eq.~\ref{eq:reduced_system}.

Previous models with asymmetric connectivity~\cite{tsodyks_associative_1995, theodoni_theta-modulation_2018} have not been able to generate reverse sequences. The belief that symmetric connectivity is necessary for bidirectional replay has been stated in many existing works~\cite{gupta_hippocampal_2010, haga_recurrent_2018, ecker_hippocampal_2022, milstein_offline_2023, mallory_time_2025}. 
Based on the large amount of apparent evidence with articles claiming that "[asymmetric distributions of synaptic weights] cannot account for reverse replay"~\cite{milstein_offline_2023}, one would be lead to believe that their belief is indeed true. As we have shown, it is not.

In a neural field (or network) with symmetric connectivity, a symmetry-breaking mechanism through a slow fatigue variable is required for bidirectional replay. The direction of a traveling wave is then decided by the spatially asymmetric state of the fatigue variable~\cite{azizi_computational_2013, ecker_hippocampal_2022, milstein_offline_2023,romani_short-term_2015, pietras_mesoscopic_2022, haga_recurrent_2018}. 
\citeauthor{mallory_time_2025}\cite{mallory_time_2025} observed self-avoidance of preceding trajectories in replays, evidence for the important role of such a "neuronal fatigue" mechanism in CA3. Biologically, it is typically attributed to spike-frequency adaptation (SFA), which acts on the order of seconds and has been observed in a large fraction of cortical neurons~\cite{salaj_spike_2021}: After repeated spiking from prolonged stimulation, a neuron's spiking frequency decreases due to an increased firing threshold.
Short-term synaptic depression, sometimes treated as a special case of SFA~\cite{ganguly_spike_2024}, describes the depletion of synaptic vesicles after repeated firing and acts as another symmetry-breaking fatigue mechanism. While our model is compatible with these mechanisms, it does not require them. Instead, our model achieves symmetry-breaking by the initialization of delayed activity (and, optionally, asymmetric weights).




\subsection{Biophysical relevance}
In this section, we inspect and discuss various simplifications and assumptions of our work. Our neural field model targets the dynamics of the hippocampal CA3 region, which is a primary source of sharp-wave ripples and replay~\cite{he_expanded_2023}. This area is characterized by high recurrence, with a local connectivity probability of approximately 10\%~\cite{sammons_structure_2024}. A key requirement of our model is the circular structure of inputs to the circuit, which aligns with biological observations: The population activity of upstream grid cells in the medial entorhinal cortex has been shown to exhibit a toroidal topology regardless of context~\cite{gardner_toroidal_2022}. Such a circular structure has also been used in previous models of hippocampal replay~\cite{romani_short-term_2015, pietras_mesoscopic_2022, li_mechanisms_2024}. Besides this periodicity of spatial representations, inputs to our model are also temporally periodic with timescale $T$. While behavioral navigation occurs over seconds, the hippocampus processes these sequences much faster. This is mediated by phase precession, where spatial trajectories are temporally compressed into individual theta cycles ($\sim$\SI{100}{\milli\second})~\cite{skaggs_theta_1996}. As a consequence, replay in our model with $|v^*| \approx 1$ occurs on this compressed timescale, which is in line with observations~\cite{denovellis_hippocampal_2021}. In this regime, the total transmission delay $\delay$ between pyramidal cells in CA3 constitutes a significant fraction of the cycle phase. This delay consists of axonal conduction with a typical velocity of \SI{\sim 350}{\micro\meter\per\milli\second}~\cite{meeks_action_2007}), synaptic latency (\SI{\sim 1.2}{\milli\second}~\cite{sammons_structure_2024}), and membrane integration time (approximated by the 20--80\% EPSP rise time: \SI{\sim 5}{\milli\second}~\cite{mishra_symmetric_2016}). Combining these measurements, we can approximate the total delay to be \SI{\sim 6}{\milli\second}. The propagation delay between two cells with a distance of \SI{\sim 500}{\micro\meter} during a sharp-wave ripple was measured to be slightly lower, between \SI{\sim 2.5}{\milli\second} and \SI{\sim 7.1}{\milli\second}~\cite{schieferstein_propagation_2024}. In either case, we can assume the delay $\delay$ to be larger than the firing rate time constant \(\tau_r\), which we set to \SI{2}{\milli\second}. This $\tau_r$ applies to pyramidal cell populations and should not be confused with the membrane time constant of a single neuron~\cite{volgushev_cortical_2016}.

Plasticity in CA3 is not purely determined by a causal learning rule, as chosen in our simulations, but likely contains both symmetric and asymmetric components depending on the novelty of the stimulus~\cite{madar_btsp_2023}. Using our model, the asymmetry of synaptic connectivity \(\text{w}\) can be derived from the ratio between forward and reverse replay speeds. If the connectivity is symmetric, then reverse replay has the same speed and occurs equally often as forward replay. Because forward replay has been observed to occur more often than reverse replay, we can deduce that the connectivity kernel for replay in CA3 has a causal component.

\subsection{Initiation and termination of replay}
Our analytical results describe a set of possible self-sustaining activity patterns but do not prescribe which pattern is selected. From a dynamical systems perspective, the neural circuit acts as a multistable system with distinct basins of attraction corresponding to forward and reverse replay modes at various discrete speeds. The biological initiation of replay can thus be understood as a state-selection process.
In vivo, the cortex can exert top-down control~\cite{mallory_time_2025} and act as the initial seed of activity. During wakefulness, sensory information biases backward replays towards starting from the animal's current position, but different replay starting points have also been found~\cite{gupta_hippocampal_2010, davidson_hippocampal_2009}. Our simulations demonstrate that a short, noisy pattern presentation is sufficient for the network to settle into the nearest stable traveling wave solution (see Fig~\ref{fig:realistic_replay}).

The fact that replay events in vivo are finite bursts rather than infinite loops suggests that while the recurrent dynamics we describe drive propagation, an additional mechanism, such as SFA, is likely required to terminate the sequence. \citeauthor{musset_microcircuit_2026} suggest that short-term synaptic depression serves as the termination mechanism, effectively transforming the stable limit cycles predicted by our reduced model into transient, metastable states~\cite{musset_microcircuit_2026}.





\section{Acknowledgements}
Lizzy Robertson for insightful discussions about potential realizations of our proposed mechanism in analog hardware.

\appendix
\section{Simulation details}
We used Tsitouras' 5/4 method~\cite{tsitouras_rungekutta_2011} with a step size of \SI{1}{{\micro\second}} for the numerical integration for Fig.~\ref{fig:bifurcation_diagram} and Figs.~\ref{fig:numerical_stability_analysis_stable}-\ref{fig:numerical_stability_analysis_unclear}. For the remaining figures, we used Euler's method with a step size of \SI{50}{{\micro\second}}. Simulation and plotting code is available at \url{https://github.com/1b15/bidirectional_hippocampal_replay}.

\textbf{Default parameter values.}
\label{sec:parametervalues}
Unless stated otherwise, all figures for our neural field model (Sec.~\ref{sec:model}) were created with the following parameters.
{
\vspace{-0.3cm}
\renewcommand{\arraystretch}{1.25}
\newcommand{\sectiontitlespace}{0mm}
\newcommand{\sectionspace}{-0.25cm}
\newcommand{\cellspace}{0.1cm}
\newcommand{\postcellspace}{0.1cm}
\vspace{-0.45cm}
\begin{longtable}{l l l}
    \multicolumn{3}{l}{} \\[\sectiontitlespace]
    Parameter & Value & Description \\
    \hline
    $N$ & 700 & Number of discrete spatial units \\
    $T$ & \SI{35}{{\milli\second}} & Period of periodic input current \\
    $\tau_r$ & \SI{2}{{\milli\second}} & Effective firing rate time constant \\
    $\delay$ & \SI{5}{{\milli\second}} & Transmission delay \\
    $\tau_{\text{w}}$ & \SI{20}{{\second}} & Synaptic plasticity time constant \\
    $c_{u}$ & 5000 & Scale factor of input current \\
    $\weightdecay$ & 50 & Weight decay factor \\
\end{longtable}
}

\textbf{FitzHugh-Nagumo parameter values.}
\label{sec:fhn_parameters}
For our FHN model in Fig.~\ref{fig:fhn}, we used the following parameters.
{
\vspace{-0.6cm}
\renewcommand{\arraystretch}{1.25}
\begin{longtable}{l l l}
    \multicolumn{3}{l}{} \\[0mm]
    Parameter & Value & Description \\
    \hline
    $N$ & 512 & Number of discrete spatial units \\
    $L$ & 100 & Spatial domain length \\
    $\tau_\text{v}$ & 1.5 & Excitation time constant \\
    $\tau_w$ & 12.5 & Recovery time constant \\
    $\delay$ & 50.0 & Transmission delay \\
    $a$ & 0.7 & \\
    $b$ & 0.8 & \\
    $\Delta t$ & 0.01 & Simulation time step \\
\end{longtable}
}

\section{Synaptic interaction kernel representation}
\label{sec:kernelrepresentation}
{
Starting from a fully connected recurrent network in the spatial continuum limit (compare Fig.~\ref{fig:basic_activity_with_weights}), we can express the firing rate dynamics as
\begin{align*}
&\tau_r {\partial_t} r(x, t) \\
&= -r(x, t) + \activation\left(u(x, t) + \int_0^T \text{w}(x, y, t) r(y, t - \delay) dy\right),
\end{align*}
and learning dynamics as
\begin{equation}\label{eq:diffHebb}
    \tau_{\text{w}} {\partial_t} \text{w}(x, y, t) = {r}(y, t-\delay){\partial_t} r(x, t) - \weightdecay \text{w}(x, y, t),
\end{equation}
where \({ \text{w}(x, y, t) }\) represents the synaptic weight from position \({ y }\) to position \({ x }\).

We simplify \({\text{w}(x, y, t)}\) into a weight kernel \({\text{w}(\Delta x, t)}\), reflecting the fact that the synaptic efficacies of our system naturally converge to the same value for neurons at a fixed phase difference.
\begin{figure}[!htbp]
  \centering
  \includegraphics[width=0.9\linewidth]{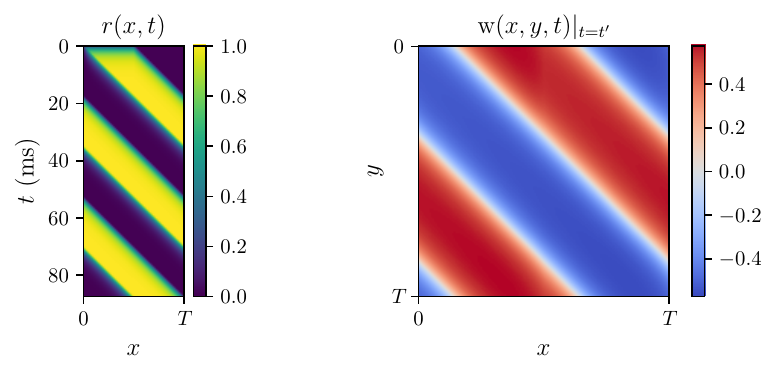}
  \vspace*{-0.4cm}
  \caption{Left: Evolution of the firing rate over space and time for a traveling sinusoidal stimulation. Right: Learned synaptic weights from \({y}\) to \({x}\) for the same stimulation. The value range of \({\text{w}(x, y, t)}\) highly depends on \({\weightdecay}\).}
  \label{fig:basic_activity_with_weights}
\end{figure}
A two-dimensional function \({ f(x, y) }\) is termed a circulant kernel if it exhibits the property
\begin{equation*}
  f(x, y) = f(x + \Delta, y + \Delta)
\end{equation*}
for any fixed shift \({ \Delta }\) within its domain (see Figure~\ref{fig:basic_activity_with_weights}). This circulant nature implies that the function values are invariant under equal shifts in both its arguments.
For a circulant kernel \({ f(x, y) }\), the integral transform \({\int f(x, y) g(y) dy}\) can be expressed as a cyclic convolution with the kernel \({ h(\Delta) = f(x, x - \Delta) }\). This results in the following equivalence:
\begin{equation}
\label{eq:circulantconv}
\int f(x, y) g(y) dy = \int h(x - y) g(y) dy
\end{equation}

Given symmetric stimulation and separation of time scales \({ \tau_{\text{w}} \gg \tau_r }\), the weights \({ \text{w}(x, y, t) }\) after long-term evolution can be approximated as circulant.
The separation of time scales allows us to approximate the evolution of synaptic weights using cycle-averaged learning dynamics. Starting from the differential Hebbian learning rule from Eq.~\eqref{eq:diffHebb}, we have
\begin{gather*}
\int_{0}^{T}\tau_{\text{w}} {\partial_t} \text{w}(x, y, t+t') dt' \nonumber \\
= \int_{0}^{T}{r}(y, t+t'-\delay){\partial_t} {r}(x, t+t')dt' \\
- \int_{0}^{T}\weightdecay \text{w}(x, y, t+t') dt'.
\end{gather*}

Exploiting the symmetry of \({ r(x, t) }\), we instead integrate over \({ x }\):
\begin{equation*}
\int_{0}^{T}{r}(y+x', t-\delay){\partial_t} {r}(x+x', t)dx' - \int_{0}^{T}\weightdecay \text{w}(x, y, t+t') dt'.
\end{equation*}

After circularly shifting the integration by \({ -y }\), the equation becomes
\begin{equation*}
\int_{0}^{T}{r}(x', t-\delay){\partial_t} {r}(x-y+x', t)dx' - \int_{0}^{T}\weightdecay \text{w}(x, y, t+t') dt'.
\end{equation*}

By assuming a slow learning rate, we approximate the cycle-averaged weight decay as the instantaneous weight decay and obtain the new learning dynamics (compare Fig.~\ref{fig:weight_kernel_evolution})
\begin{align}
& \tau_{\text{w}}{\partial_t} \text{w}(\Delta x, t)\nonumber\\
& = \frac{1}{T} \int_0^T {r}(x', t-\delay){\partial_t} {r}(\Delta x + x', t)dx' - \weightdecay \text{w}(\Delta x, t)\nonumber\\
& = \frac{1}{T}\left[r_{\text{pre}} \boldsymbol{\star} {\partial_t} {r}\right](\Delta x, t) - \weightdecay \text{w}(\Delta x, t),\label{eq:learningdynamics}
\end{align}
where we have used the cross-correlation operator~\({ \boldsymbol{\star} }\).

Using the convolution property of circulant kernels (Eq.~\eqref{eq:circulantconv}), the rate dynamics equation becomes
\begin{align}
&\tau_r {\partial_t} r(x, t)\nonumber\\ 
& = -r(x, t) + \activation\left(u(x, t) + \int_0^T \text{w}(x', t) r_{\text{pre}}(x - x', t) dx'\right)\nonumber\\
& = -r(x, t) + \activation\left(u(x, t) + \left[\text{w} * r_{\text{pre}}\right](x, t)\right).\label{eq:ratedynamics}
\end{align}
An empirical verification of the approximation can be seen in Figure~\ref{fig:weight_kernel_estimate}.
\begin{figure}[!htbp]
  \centering
  \includegraphics[width=0.9\linewidth]{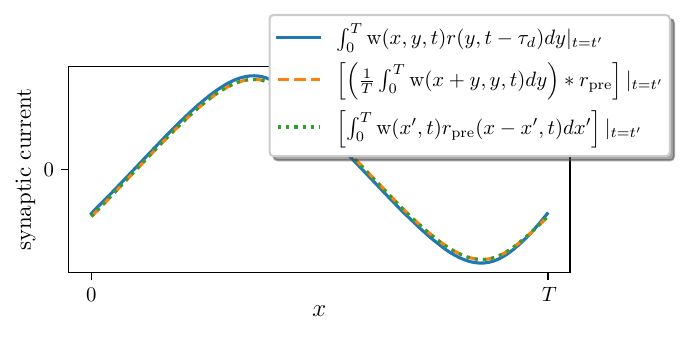}
  \vspace*{-0.3cm}
  \caption{Using the kernel weight representation, synaptic currents can be accurately approximated. The blue plot shows the recurrent signal at time $t'$, as computed by a learned fully connected synaptic weight kernel. The dashed orange plot shows the same signal, as computed by cycle-averaging the weight kernel and convolving with $r_{\text{pre}}$. In the dotted green plot, we already used the reduced representation during learning of the weight kernel.}
  \label{fig:weight_kernel_estimate}
\end{figure}
}

\section{Cubic approximation of replay speed}
\label{sec:cubicapproximation}
\begin{figure}[ht!]
  \centering
  \includegraphics[width=0.8\linewidth]{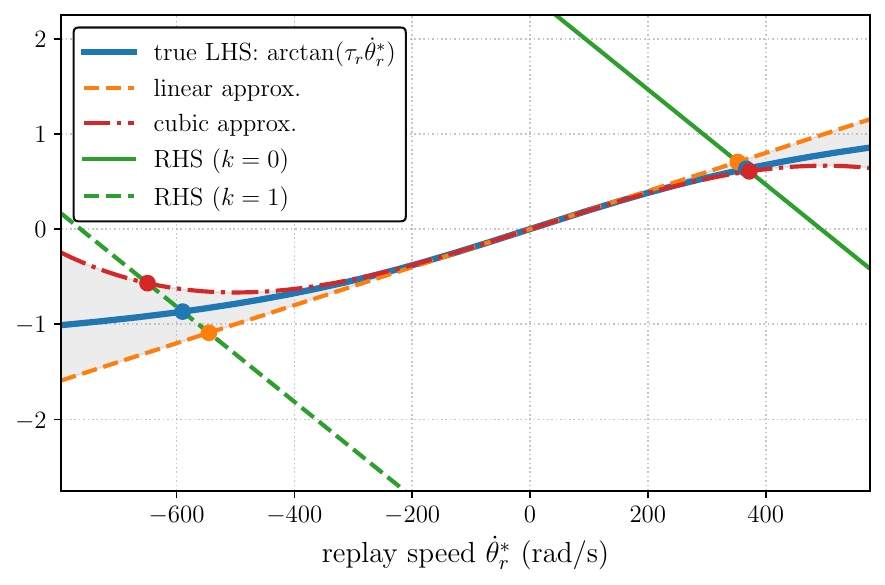}
  \caption{Linear (orange) and cubic (red) Taylor approximations of $\arctan(\tau_r\omega^*)$ (blue) provide lower and upper bounds for $\omega^*$. Green plots show the right-hand side of Eq.~\ref{eq:arctan} for $k=0$ and $k=1$, with crossings marking fixed replay speeds $\omega^*$ and their approximations.}
  \label{fig:arctan_taylor_approximation}
\end{figure}
This section derives a third-order Taylor approximation for Eq.~\ref{eq:fixed_phase_velocity}:
{
\allowdisplaybreaks
\begin{gather}
    \tau_r \omega^* = \tan\left(-\theta_{\text{w}} - \delay\omega^* - 2\pi\cyclecount - \varepsilon^* \right) \nonumber \\
    \arctan\left(\tau_r \omega^*\right) = -\theta_{\text{w}} - \delay\omega^* - 2\pi\cyclecount - \varepsilon^* \label{eq:arctan} \\
{\tau_r\omega^*} - \frac{\tau_r^3}{3} \left(\omega^*\right)^3 \approx -\theta_{\text{w}} - {\delay} \omega^* - 2\pi \cyclecount - \varepsilon^* \nonumber \\
0 = \frac{\tau_r^3}{3} \left(\omega^*\right)^3 - (\delay + \tau_r)\omega^* - \theta_{\text{w}} - 2\pi \cyclecount - \varepsilon^* \nonumber
\end{gather}
}
We divide by $\frac{\tau_r^3}{3}$ to obtain the depressed cubic
\begin{equation*}
0 = \left(\omega^*\right)^3 - \frac{3(\delay + \tau_r)}{\tau_r^3}\omega^* - \frac{3}{\tau_r^3}\left(\theta_{\text{w}} + 2\pi \cyclecount + \varepsilon^*\right)
\end{equation*}

From here, we get the trigonometric solution for three real roots
\begin{align*}
 & \omega^*=2\,{\sqrt {-{\frac {p}{3}}}}\,\cos \left[\,{\frac {1}{3}}\arccos \left({\frac {3q}{2p}}{\sqrt {\frac {-3}{p}}}\,\right)-{\frac {2\pi j}{3}}\,\right]\\
 & {\text{with }}j \in \{0,1,2\},\\
 & p = -\frac{3(\delay + \tau_r)}{\tau_r^3} \text{,\hspace{1em}} q = -\frac{3}{\tau_r^3}\left(\theta_{\text{w}} + 2\pi \cyclecount + \varepsilon^*\right)
\end{align*}
We are only interested in the case $j=1$.

\section{Standing wave solutions}
\begin{figure}[ht!]
  \centering
  \includegraphics[width=0.72\linewidth]{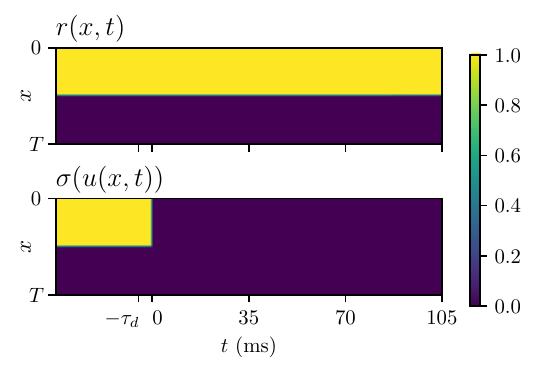}
  \vspace*{-0.3cm}
  \caption{Replay of standing wave, resulting from a weight kernel learned with \({\delay = \frac{3}{4}T}\).}
  \label{fig:simple_standing_wave}
\end{figure}

In Eq.~\ref{eq:fixed_phase_velocity}, we computed fixed replay speeds from biophysical parameters. Now we consider the special case of standing waves, \({\omega^* = 0}\), and compute the corresponding parameters.
From Eq.~\ref{eq:linearapprox_weightphase} it follows that 
\begin{equation*}
  \omega^* = 0 \Leftrightarrow -\theta_{\text{w}} - 2\pi \cyclecount - \varepsilon \approx 0.
\end{equation*}
For a weight kernel that was obtained through differential Hebbian learning, we get from Eq.~\ref{eq:linearapprox}:
\begin{equation*}
  \omega = 0 \Leftrightarrow \cyclecount + \frac{1}{4} + \frac{\delay}{T} - \frac{\varepsilon}{2\pi} \approx 0\\
\end{equation*}
Assuming \({\delay \in (0, T)}\), there is a unique solution for delays that produce standing waves: \({\delay \approx \frac{3}{4}T}\) (as shown in Fig.~\ref{fig:simple_standing_wave}).

\section{Replay dynamics in the co-rotating coordinate frame}
\label{sec:rotating_coordinate_frame_derivation}
To express the system in a frame where the fundamental phase is stationary, we substitute the relative phase $\vartheta(t)$ from Eq.~\eqref{eq:vartheta} into the governing dynamics, Eq.~\eqref{eq:reduced_system}. The time derivative transforms as $\dot{\theta}_r(t) = \dot{\vartheta}(t) - \omega^*$, and the delayed phase difference expands to
\begin{equation}
    \theta_r(t-\delay) - \theta_r(t) = \vartheta(t-\delay) - \vartheta(t) + \omega^* \delay.
\end{equation}
Substituting this into the trigonometric argument in Eq.~\eqref{eq:reduced_system} yields
\begin{equation*}
    \theta_{\text{w}} - \omega^* \delay + \varepsilon + \vartheta(t-\delay) - \vartheta(t).
\end{equation*}
Exploiting the $2\pi$-periodicity of the sine and cosine functions, we absorb the constant terms and an arbitrary winding number $\cyclecount \in \mathbb{Z}$ into a single phase offset, $\phi^* = \theta_{\text{w}} + \omega^* \delay + 2\pi\cyclecount + \varepsilon^*$.

Replacing the argument in Eq.~\eqref{eq:amp_eq} with ${\phi^* + \vartheta(t-\delay) - \vartheta(t)}$ immediately yields the rotating-frame amplitude dynamics [Eq.~\eqref{eq:amp_dot_rotating_frame}]. Similarly, substituting the argument and $\dot{\theta}_r(t)$ into Eq.~\eqref{eq:phase_eq}, and subsequently dividing by $\tau_r a_r(t)$, yields the rotating-frame phase dynamics [Eq.~\eqref{eq:vartheta_dot_rotating_frame}].

\section{Derivation of the Jacobian matrices for linear stability analysis}
\label{sec:stability_derivation}
Let the vector field of the system described by Eqs.~\eqref{eq:amp_dot_rotating_frame} and \eqref{eq:vartheta_dot_rotating_frame} be denoted by
\begin{align*}
    &F_a\left(a_r(t), \vartheta(t), \vartheta(t-\delay)\right) \nonumber\\
    &\quad = -a_r(t) + \frac{2}{\pi} \cos\left(\phi^* +\vartheta(t-\delay) - \vartheta(t)\right), \\
    &F_\vartheta\left(a_r(t), \vartheta(t), \vartheta(t-\delay)\right) \nonumber\\
    &\quad = \frac{1}{a_r(t)} \frac{2}{\pi} \sin\left(\phi^* +\vartheta(t-\delay) - \vartheta(t) \right) - \tau_r \omega^*.
\end{align*}
The elements of the Jacobian matrices are the partial derivatives of this vector field evaluated at the fixed point $(a_r^*, \vartheta^*)$.
\begin{equation*}
    [\mathbf{M}_0]_{11} = \left.\frac{\partial F_a}{\partial a_r(t)}\right|_* = \left.\frac{\partial}{\partial a_r(t)}\left[-a_r(t)\right]\right|_* = -1.
\end{equation*}
\begin{align*}
    [\mathbf{M}_0]_{12} &= \left.\frac{\partial F_a}{\partial \vartheta(t)}\right|_* \nonumber\\
    &= \left.\frac{\partial}{\partial \vartheta(t)}\left[\frac{2}{\pi} \cos\left(\phi^* +\vartheta(t-\delay) - \vartheta(t)\right)\right]\right|_* \nonumber\\
    &= \frac{2}{\pi} \sin(\phi^*)\nonumber
    \intertext{Using the fixed-point condition from Eq.~\eqref{eq:phase_eq}, ${\tau_r a_r^*\omega^* = \frac{2}{\pi}\sin(\phi^*)}$, this simplifies to:}
    [\mathbf{M}_0]_{12} &= -a_r^* \tau_r \omega^*.
\end{align*}
The amplitude dynamics do not depend on the delayed amplitude, so
\begin{equation*}
    [\mathbf{M}_{\delay}]_{11} = \left.\frac{\partial F_a}{\partial a_r(t-\delay)}\right|_* = 0.
\end{equation*}
\begin{align*}
    [\mathbf{M}_{\delay}]_{12} &= \left.\frac{\partial F_a}{\partial \vartheta(t-\delay)}\right|_* \nonumber\\
    &= \left.\frac{\partial}{\partial \vartheta(t-\delay)}\left[\frac{2}{\pi} \cos\left(\phi^* +\vartheta(t-\delay) - \vartheta(t)\right)\right]\right|_* \nonumber\\
    &= -\frac{2}{\pi} \sin(\phi^*) = a_r^* \tau_r \omega^*.
\end{align*}

The phase perturbation dynamics are:
\begin{align*}
    [\mathbf{M}_0]_{21} &= \left.\frac{\partial F_\vartheta}{\partial a_r(t)}\right|_* = \left.\frac{\partial}{\partial a_r(t)}\left[\frac{1}{a_r(t)} \frac{2}{\pi} \sin\left(\dots \right)\right]\right|_* \nonumber\\
    &= -\frac{1}{(a_r^*)^2} \frac{2}{\pi} \sin(\phi^*) \nonumber\\
    &= -\frac{1}{(a_r^*)^2} (\tau_r a_r^* \omega^*) = \frac{\tau_r \omega^*}{a_r^*}.
\end{align*}
\begin{align*}
    [\mathbf{M}_0]_{22} &= \left.\frac{\partial F_\vartheta}{\partial \vartheta(t)}\right|_* \nonumber\\
    &= \left.\frac{\partial}{\partial \vartheta(t)}\left[\frac{1}{a_r(t)} \frac{2}{\pi} \sin\left(\phi^* +\vartheta(t-\delay) - \vartheta(t) \right)\right]\right|_* \nonumber\\
    &= -\frac{1}{a_r^*} \frac{2}{\pi} \cos(\phi^*)\nonumber\\
    \intertext{Using the amplitude fixed-point condition from Eq.~\eqref{eq:amp_eq}, $a_r^* = \frac{2}{\pi}\cos(\phi^*)$, this becomes:}
    [\mathbf{M}_0]_{22} &= -1.
\end{align*}
The phase dynamics do not depend on the delayed amplitude, so
\begin{equation*}
    [\mathbf{M}_{\delay}]_{21} = \left.\frac{\partial F_\vartheta}{\partial a_r(t-\delay)}\right|_* = 0.
\end{equation*}
\begin{align*}
    &[\mathbf{M}_{\delay}]_{22} = \left.\frac{\partial F_\vartheta}{\partial \vartheta(t-\delay)}\right|_* \nonumber\\
    &= \left.\frac{\partial}{\partial \vartheta(t-\delay)}\left[\frac{1}{a_r(t)} \frac{2}{\pi} \sin\left(\phi^* +\vartheta(t-\delay) - \vartheta(t) \right)\right]\right|_* \nonumber\\
    &= \frac{1}{a_r^*} \frac{2}{\pi} \cos(\phi^*) = 1.
\end{align*}

\section{Derivation of the Hopf bifurcation condition}
\label{sec:Hopf_derivation}
A Hopf bifurcation occurs when a pair of complex conjugate eigenvalues crosses the imaginary axis. We therefore probe the boundary of stability by setting $\lambda = i\Omega$, where $\Omega \in \mathbb{R}$ and $\Omega > 0$ is the angular frequency of the emerging oscillation. Substituting this into the characteristic equation (Eq.~\eqref{eq:characteristic_equation}) yields
\begin{equation*}
    (\tau_r i\Omega + 1)^2 + (\tau_r \omega^*)^2 = \left[(\tau_r i\Omega + 1) + (\tau_r \omega^*)^2\right] e^{-i\Omega \delay}.
\end{equation*}
Expanding the complex terms, we can separate the real and imaginary parts:
\begin{align*}
    &\left[1 - (\tau_r \Omega)^2 + (\tau_r \omega^*)^2\right] + i\left[2\tau_r\Omega\right] \\
    &\quad= \left[ (1+(\tau_r \omega^*)^2) + i\tau_r\Omega \right] e^{-i\Omega \delay}.
\end{align*}
For this equality to hold, the squared magnitudes of the complex numbers on each side must be equal. As $|e^{-i\Omega \delay}|^2 = 1$, this requires
\begin{align*}
    &\left(1+(\tau_r \omega^*)^2 - (\tau_r \Omega)^2\right)^2 + (2\tau_r\Omega)^2 \\
    &\quad = \left(1+(\tau_r \omega^*)^2\right)^2 + (\tau_r\Omega)^2.
\end{align*}
Expanding the squared terms, the common term ${(1+(\tau_r \omega^*)^2)^2}$ cancels, leading to
\begin{equation*}
    -2\left(1+(\tau_r \omega^*)^2\right)(\tau_r \Omega)^2 + (\tau_r \Omega)^4 + 4(\tau_r\Omega)^2 = (\tau_r\Omega)^2.
\end{equation*}
Assuming a non-trivial oscillation frequency ($\Omega \neq 0$), we can divide the entire equation by $(\tau_r \Omega)^2$ and rearrange to solve for the squared Hopf frequency:
\begin{align}
    -2\left(1+(\tau_r \omega^*)^2\right) + (\tau_r \Omega)^2 + 4 &= 1, \nonumber \\
    (\tau_r \Omega)^2 &= 2(\tau_r \omega^*)^2 - 1. \label{eq:appendix_Hopf_frequency_sq}
\end{align}
A Hopf bifurcation occurs at a real, non-zero frequency $\Omega$, which requires $\Omega^2 > 0$. From Eq.~\eqref{eq:appendix_Hopf_frequency_sq}, this restricts the existence of a bifurcation to the regime:
\begin{equation*}
    (\tau_r\omega^*)^2 > \frac{1}{2}.
\end{equation*}
To determine the stability for $(\tau_r\omega^*)^2 \le 1/2$, we can evaluate the characteristic equation at the limit $(\tau_r\omega^*)^2 = 0$, which reduces to
\begin{equation*}
    (\tau_r \lambda + 1)^2 = (\tau_r \lambda + 1) e^{-\lambda \delay}.
\end{equation*}
The non-trivial roots of this equation are $\lambda = -1/\tau_r$ and solutions to $\tau_r \lambda + 1 = e^{-\lambda \delay}$. To see that the latter equation only has solutions with $\text{Re}(\lambda) \le 0$, we let $\lambda = \mu + i \Omega$ and equate the absolute magnitudes of both sides:
\begin{equation*}
    \sqrt{(\tau_r \mu + 1)^2 + (\tau_r \Omega)^2} = e^{-\mu \delay}.
\end{equation*}
If we assume an unstable root exists, such that $\mu > 0$ (and given $\tau_r > 0, \delay > 0$), the left-hand side is strictly greater than 1, while the right-hand side is strictly less than 1. This is a contradiction; therefore, no solutions with $\mu > 0$ can exist, meaning all roots are restricted to $\text{Re}(\lambda) \le 0$. Because eigenvalues vary continuously with system parameters, and no purely imaginary roots can exist to allow a bifurcation for $(\tau_r\omega^*)^2 \le 1/2$, the system must remain stable throughout this entire region. Therefore, $(\tau_r\omega^*)^2 > 1/2$ is a strict lower bound for the onset of this oscillatory instability.

\section{Numerical stability analysis}
\label{sec:numerical_stability_analysis}
\begin{figure}[ht!]
  \centering
  \includegraphics[width=0.95\linewidth]{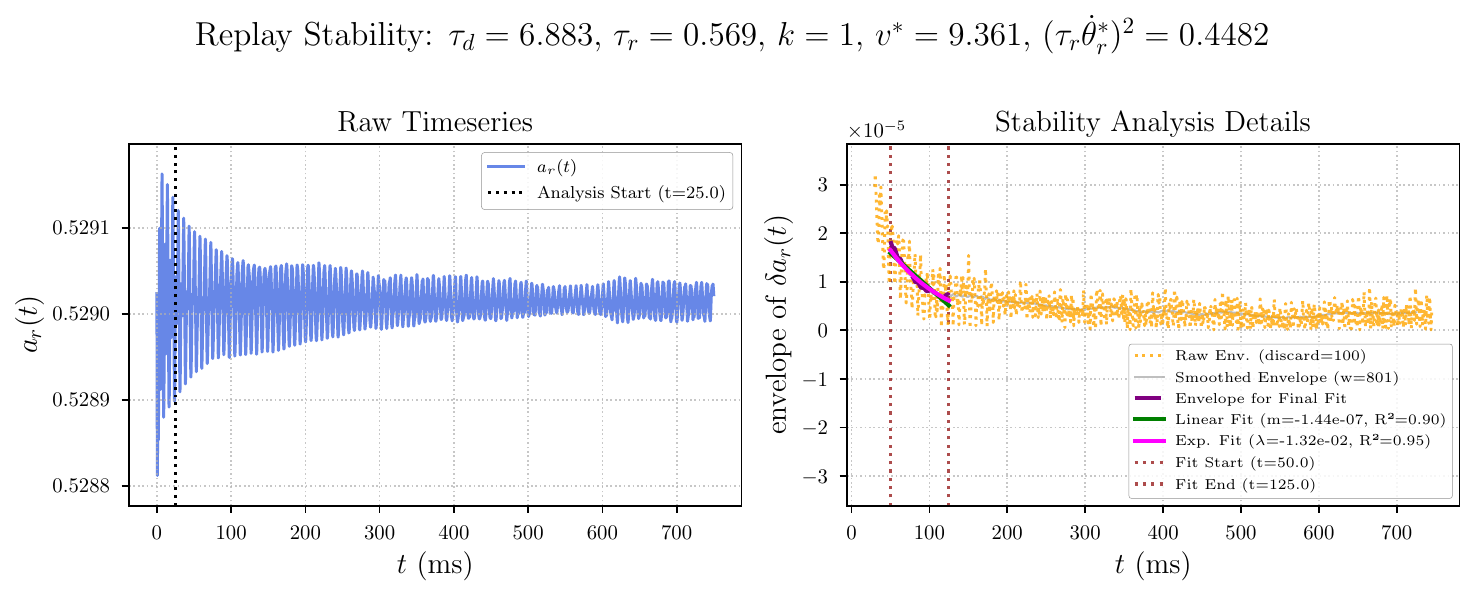}
  \caption{Example of stable replay solution.}
  \label{fig:numerical_stability_analysis_stable}
\end{figure}

\begin{figure}[ht!]
  \centering
  \includegraphics[width=0.95\linewidth]{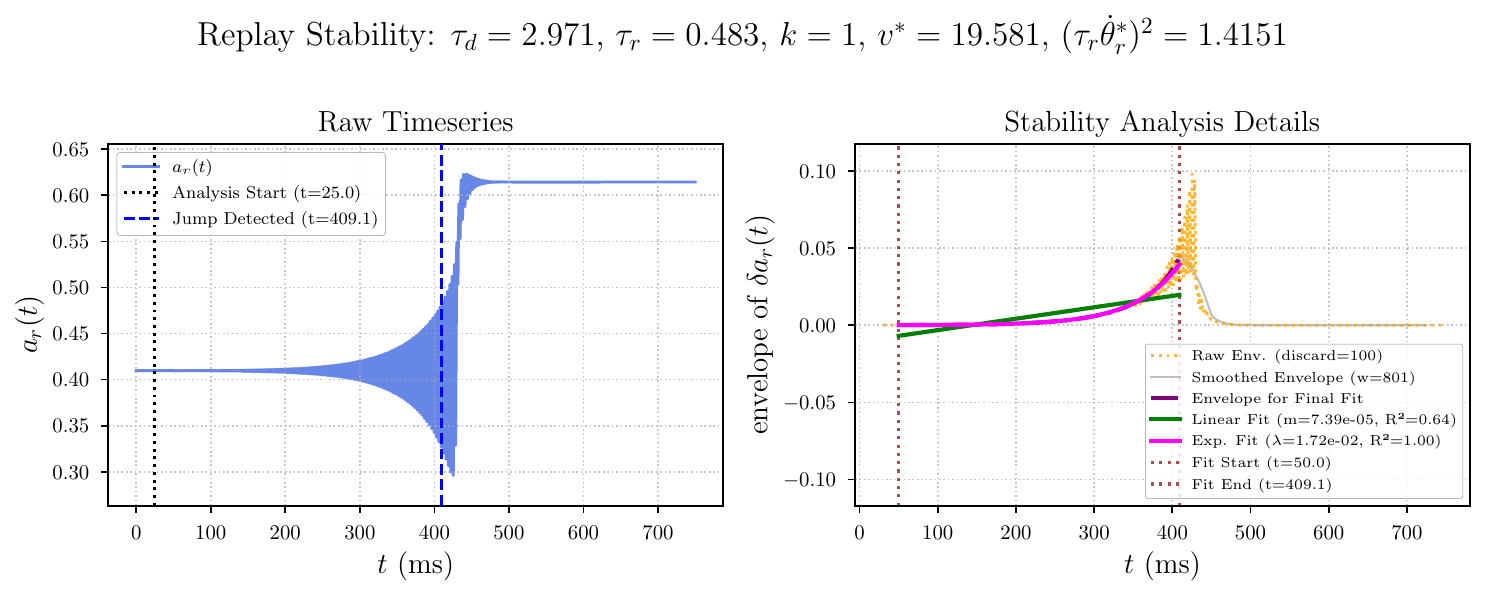}
  \caption{Example of unstable replay solution. The exponential is only fitted until a jump to a higher amplitude / lower absolute replay speed is detected.}
  \label{fig:numerical_stability_analysis_unstable}
\end{figure}

\begin{figure}[ht!]
  \centering
  \includegraphics[width=0.95\linewidth]{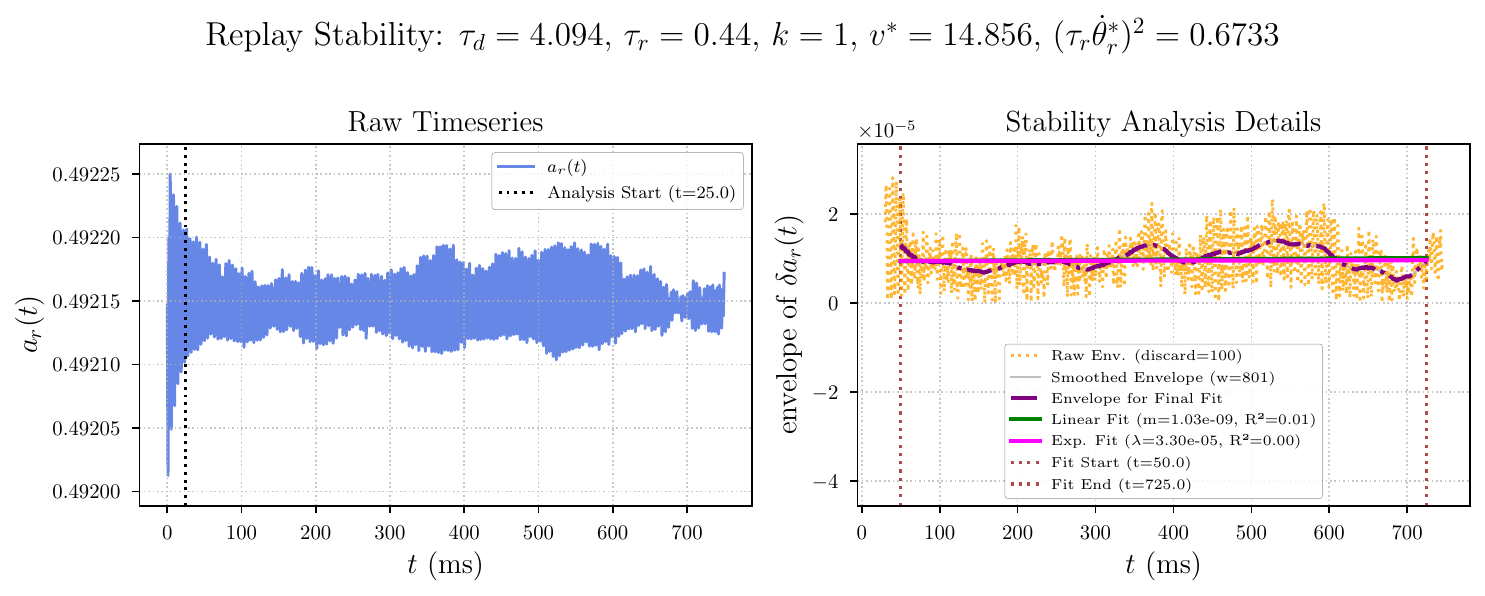}
  \caption{Example of replay solution with ambiguous stability.}
  \label{fig:numerical_stability_analysis_unclear}
\end{figure}

In order to numerically determine the stability of simulated replay solutions, we fit an exponential function to the smoothed envelope of \(\delta a_r(t)\). Stable solutions provide a good fit (\(R^2 > 0.8\)) with negative exponent (Fig.~\ref{fig:numerical_stability_analysis_stable}), and unstable solutions a good fit with positive exponent (Fig.~\ref{fig:numerical_stability_analysis_unstable}). In certain cases (Fig.~\ref{fig:numerical_stability_analysis_unclear}), the exponential function can't be accurately fitted and we don't make a statement about the solution's stability.

\clearpage
\bibliography{references}


\end{document}